\documentclass[journal,twoside]{IEEEtran}
\usepackage{amssymb}
\usepackage{amsfonts}
\usepackage{amsmath}
\usepackage{graphicx}
\usepackage{color}

\begin{document}

\title{Design of Ultra-compact Graphene-based Superscatterers}
\author{Rujiang Li, Bin Zheng, Xiao Lin, Ran Hao, Shisheng Lin, Wenyan Yin,~\IEEEmembership{Fellow,~IEEE},
Erping Li,~\IEEEmembership{Fellow,~IEEE}, and Hongsheng Chen
\thanks{%
Manuscript received August XX, 2015; revised XXXXXX XX, 201X. This work was sponsored
by the National Natural Science Foundation of China under Grants No. 61322501,
No. 61574127, and No. 61275183, the National Program for Special Support of
Top-Notch Young Professionals, the Program for New Century Excellent Talents
(NCET-12-0489) in University, the Fundamental Research Funds for the Central
Universities, and the Innovation Joint Research Center for Cyber-Physical-Society
System.}
\thanks{
R. Li, B. Zheng, X. Lin, and H. Chen are with the State Key Laboratory of
Modern Optical Instrumentation, Zhejiang University, Hangzhou 310027, China,
with the College of Information Science and Electronic Engineering, Zhejiang
University, Hangzhou 310027, China, and also with The Electromagnetics
Academy of Zhejiang University, Zhejiang University, Hangzhou 310027, China
(e-mails: B. Zheng, zhengbin@zju.edu.cn; H. Chen, hansomchen@zju.edu.cn).}
\thanks{
R. Hao, S. Lin, W. Yin, E. Li is with the College of Information Science and Electronic Engineering, Zhejiang
University, Hangzhou 310027, China (e-mail: E. Li, liep@zju.edu.cn).}
\thanks{%
Color versions of one or more of the figures in this paper are available
online at http://ieeexplore.ieee.org.} \thanks{%
Digital Object Identifier}}
\maketitle

\begin{abstract}
The energy-momentum dispersion relation is a fundamental
property of plasmonic systems. In this paper,
we show that the method of dispersion engineering can be used
for the design of ultra-compact graphene-based superscatterers.
Based on the Bohr model, the dispersion relation of the equivalent planar
waveguide is engineered to enhance the scattering cross section of a dielectric
cylinder. Bohr conditions with different orders are fulfilled
in multiple dispersion curves at the same resonant frequency.
Thus the resonance peaks from the
first and second order scattering
terms are overlapped in the deep-subwavelength scale
by delicately tuning the gap thickness
between two graphene layers. Using this ultra-compact graphene-based superscatterer,
the scattering cross section of the dielectric cylinder can be enhanced
by five orders of magnitude.

\end{abstract}

\markboth{IEEE JOURNAL OF SELECTED TOPICS IN QUANTUM ELECTRONICS,~Vol.~XX, No.~X, XXXXXXXXX~201X}
{Li \MakeLowercase{\textit{et al.}}: Design of Ultra-compact Graphene-based Superscatterers}


\IEEEpeerreviewmaketitle

\section{Introduction}

\IEEEPARstart{S}{uperscatterer} is a device that can magnify the scattering
cross section of a given object remarkably \textcolor[rgb]{0,0,0}{\cite{OE16-18545,PIER}}.
This concept was first proposed based on the
transformation optics approach, where the scattering cross section of a
cylindrical perfect electric conductor (PEC) is enhanced by an
anisotropic and inhomogeneous
electromagnetic cover \textcolor[rgb]{0,0,0}{\cite{OE16-18545,PIER,
NJP10-113016,APL94-223513,NJP11-073033,
OE18-6891,CMS49-820,JOSAA30-1698}}.
Alternatively, for subwavelength
objects, a superscatterer can be designed by the metal-dielectric layers,
where the scattering cross section is magnified due to the
enhancement of the localized surface plasmons \cite{PRL90-057401,JPCC}.
Besides, the concept of ``surface superscatterers'' is also proposed,
where the scattering cross section of a deep-subwavelength dielectric
cylinder is enhanced by a monolayer graphene sheet which is only one atom thick
\cite{OL40-1651,nanotechnology}.
Moreover, the scattering cross sections of subwavelength superscatterers can be
further enhanced by overlapping the resonances from different scattering terms
\cite{PRL105-013901,APL98-043101,
prl108-083902,PRA86-033825,OE21-10454,JPCC118-30170,nanoscale6-9093,APL105-011109}.
Since compact superscatterers are more promising in the miniaturization and integration
of plasmonic devices, it is quite necessary to design superscatterers by
overlapping different resonance peaks in the deep-subwavelength scale.

Besides, inspired by the Bohr quantization condition in quantum mechanics,
a Bohr model that relates the localized surface plasmons and propagating surface plasmons
has been proposed recently \cite{sr5-12148}. Based on this
geometric interpretation, the superscattering phenomenon can be understood
intuitively by the dispersion relation of the equivalent one dimensional planar
plasmonic waveguide.
Thus it is feasible to tune the scattering cross sections of subwavelength objects
and design the corresponding superscatterers by dispersion engineering.
Since the energy-momentum dispersion relation is a fundamental property
of photonic and plasmonic systems, dispersion engineering has been
used to realize various intriguing phenomena in photonic crystals and plasmonic
crystals, e.g.
slow light \cite{nphoton2-465}, spontaneous emission \cite{nphoton1-449},
trapped plasmons \cite{PRL95-116802}, and
plasmon-induced transparency \cite{PRL101-047401}.

In this paper, we show that the method of dispersion engineering
can be used to design the ultra-compact graphene-based superscatterers,
where the sizes of the superscatterers are in the deep-subwavelength
scale.
By delicately tuning the dispersion relation of the equivalent planar
waveguide, the resonance peaks from the first and second order scattering
terms are overlapped. Although the optical loss of graphene exists,
the scattering cross section of our ultra-compact superscatterer approaches the single
channel limit with the enhancement of five orders of magnitude.

This paper is organized as follows.
In section II, the applicability of Bohr model to graphene-based
structures is shown by taking the dielectric-graphene-air
cylindrical structure as an example. In section III,
based on the validation of Bohr model when multiple dispersion
curves exist simultaneously, the resonance peaks from the first and second
order scattering terms are overlapped by engineering the
dispersion relation of the equivalent planar waveguide.
In section IV, an ultra-compact graphene-based superscatterer is designed
by dispersion engineering. Finally, section V is the conclusion.

\section{Bohr model for graphene based structures}

For the scattering of plasmonic structures, the scattering models can be
related to their equivalent one dimensional planar waveguide models by Bohr
model \cite{sr5-12148}. According to the Bohr model, if the phase
accumulation along an enclosed optical path is an integral number of $2\pi $%
, namely if the Bohr condition%
\begin{equation}
\oint \beta dl=n\cdot 2\pi  \label{Bohr condition}
\end{equation}%
is satisfied, the scattering cross sections of the plasmonic structures
exhibit resonances. At the resonant frequencies, the scattering cross
sections are significantly enhanced which demonstrate the occurrences of
superscattering phenomena \cite{OL40-1651}. In Eq. (\ref{Bohr condition}),
the integer $n$ is the order of resonance, and $\beta $ is the corresponding
propagation constant of the plasmonic mode in equivalent planar waveguide.
The integral (or the enclosed optical path) is calculated along the
effective circumference of the plasmonic structures, since surface plasmons
propagate along the surface between dielectric and plasmonic material with
evanescent fields in the perpendicular directions. Specially, if the
plasmonic structure is a cylinder or sphere, Bohr condition reduces to
\begin{equation}
\beta R_{\text{eff}}=n,  \label{reduced Bohr condition}
\end{equation}%
where $R_{\text{eff}}$ is the effective radius of the cylindrical or
spherical plasmonic structures.
\textcolor[rgb]{0,0,0}{It is worth to note that,
since the plasmonic field is highly localized
on the interface and the
penetration depth is usually small compared with the radius of
the cylinder or sphere,
the propagation constant $\beta$ of surface plasmons in a curved circumference
is approximately
equal to that in an equivalent planar waveguide.
Meanwhile, the Bohr model is a phenomenological model and the
effective radius is usually determined empirically.}

Bohr model has been used to interpret the localized surface plasmons
supported by layered metal-dielectric structures \cite{sr5-12148}. Since
graphene is also one kind of plasmonic materials, in this section we show
that different orders of superscattering supported by graphene-based
structures can be interpreted by the Bohr model as well.
Besides, due to the high confinement of graphene plasmons and the one atom
thick graphene monolayer, graphene provides a suitable alternative
to metal to design the ultra-compact superscatterers in the deep-subwavelength
scale \cite{NL8-3370}.
In the calculation,
graphene layer is modeled as a two dimensional conducting film \cite{RMP}. The surface
conductivity of graphene monolayer is calculated by the Kubo formula%
\begin{equation}
\sigma _{g}\left( \omega ,\mu _{c},\tau ,T\right) =\sigma _{\text{intra}%
}+\sigma _{\text{inter}}\,,  \label{sigma_g}
\end{equation}%
where
\begin{equation}
\sigma _{\text{intra}}=\frac{ie^{2}k_{B}T}{\pi \hbar ^{2}\left( \omega
+i\tau^{-1} \right) }\left[ \frac{\mu _{c}}{k_{B}T}+2\ln \left( e^{-\mu
_{c}/k_{B}T}+1\right) \right]  \label{sigma_intra}
\end{equation}%
is due to intraband contribution, and
\begin{equation}
\sigma _{\text{inter}}=\frac{ie^{2}\left( \omega +i\tau^{-1} \right) }{\pi
\hbar ^{2}}\int_{0}^{\infty }\frac{f_{d}\left( -\varepsilon \right)
-f_{d}\left( \varepsilon \right) }{\left( \omega +i\tau^{-1} \right)
^{2}-4\left( \varepsilon /\hbar \right) ^{2}}d\varepsilon
\label{sigma_inter}
\end{equation}%
is due to interband contribution \cite{JAP103-064302,JP19-026222}. In the
above formula, $-e$ is the charge of an electron, $\hbar =h/2\pi $ is the
reduced Plank's constant, $T$ is the temperature, $\mu _{c}$ is the chemical
potential, $\tau=\mu \mu_{c}/e v_{F}^{2} $ is the carrier relaxation time,
$\mu$ is the carrier mobility which ranges from 1 000
$\text{cm}^{2}/(\text{V}\cdot\text{s})$
to 230 000 $\text{cm}^{2}/(\text{V}\cdot\text{s})$ \cite{ACSnano},
$v_{F}=c/300$ is the
Fermi velocity, $f_{d}\left(
\varepsilon \right) =1/\left[ e^{\left( \varepsilon -\mu _{c}\right)
/k_{B}T}+1\right] $ is the Fermi-Dirac distribution, and $k_{B}$ is the
Boltzmann's constant.

\begin{figure}[tbp]
\centering
\vspace{0.1cm} \includegraphics[width=8.5cm]{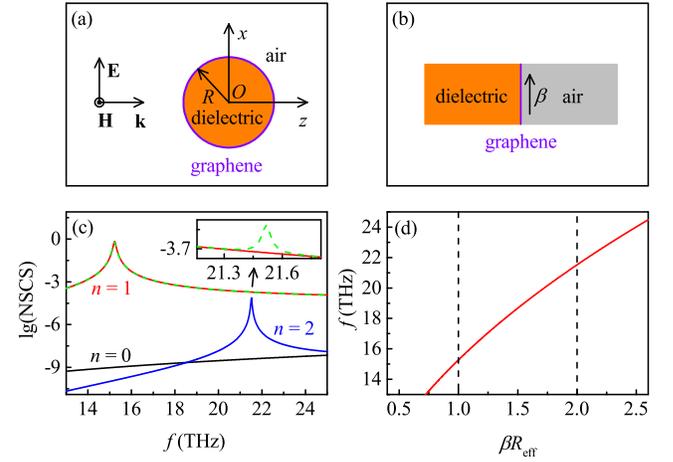} \vspace{-0.0cm}
\caption{(Color Online) (a) Cross-sectional view of the layered dielectric-
graphene-air cylindrical structure. The dielectric layers are denoted by the
orange areas, and the graphene layer is denoted by the violet area. A $x$%
-polarized plane wave is incident from air onto the structure. The radius of
the cylindrical graphene layer is $R=250$ nm. (b) Structure of the
equivalent planar waveguide model, where the graphene layer (violet area) is
separated by a semi-infinite dielectric medium (orange area) and the air
(light gray area). The graphene plasmons propagate along the graphene
surface in the direction indicated by an arrow, and $\protect\beta$ is the
corresponding propagation constant. (c) The normalized scattering cross
sections (NSCSs) at different frequencies for the structure shown in (a),
where the dashed green line denotes the total NSCS, while the solid black,
red, and blue lines denote contributions from $n=0$, $n=1$, and $n=2$
scattering terms, respectively. The inset is the enlarged figure. Note the
NSCSs are expressed in the common logarithmic form. (d) Dispersion relation
for the waveguide structure shown in (b). The two vertical dashed black
lines indicate the first ($\protect\beta R_{\text{eff}}=1$) and second ($%
\protect\beta R_{\text{eff}}=2$) order Bohr conditions, respectively. The
parameters are $\protect\varepsilon_{r}=1$, $\protect\mu_{r}=1$, $\protect\mu%
_c=0.35$ eV, $\textcolor[rgb]{0,0,0}{\mu}=$ 85 600
$\text{cm}^{2}/(\text{V}\cdot\text{s})$,
and $T=300$ K for (a)-(d).}
\label{fig1}
\end{figure}

To demonstrate the applicability of Bohr model to graphene-based structures,
we start with a simple case. As shown in Fig. \ref{fig1}(a), a $x$-polarized
plane wave is incident normally from air onto an infinite long graphene
coated dielectric cylinder. For simplicity, the relative permittivity of the
dielectric cylinder is assumed to be $\varepsilon_r=1$, and the relative
permeability is $\mu_r=1$. The radius of the cylindrical graphene layer is $%
R=250$ nm. Besides, the parameters of the graphene layer are $\mu_c=0.35 $
eV, $\textcolor[rgb]{0,0,0}{\mu}=$ 85 600
$\text{cm}^{2}/(\text{V}\cdot\text{s})$, and $T=300$ K.
According to the Bohr model, this
scattering structure has its equivalent planar waveguide structure as shown
in Fig. \ref{fig1}(b), where the graphene layer is separated by a
semi-infinite dielectric medium and the air. This planar waveguide supports
the propagation of transverse magnetic (TM) surface plasmons, and $\beta$ is
the propagation constant. From Bohr model, the superscattering phenomenon
can be understood intuitively by Figs. \ref{fig1}(a) and (b). The incident $%
x $-polarized plane wave excites different orders of TM
whispering-gallery-like modes along the graphene surface \cite{PRL105-013901}%
, where the modes can be approximated by the TM plasmonic modes in
equivalent planar waveguide since the penetration depth of graphene plasmon
is smaller than the radius of the cylindrical graphene layer. When one or
more whispering-gallery-like modes satisfy the Bohr condition (similar to
the whispering gallery condition in Refs. \cite{PRL105-013901,APL98-043101}%
), the localized electromagnetic fields interfere constructively, and the
scattering cross sections are enhanced with resonance peaks accordingly.

For the scattering model in Fig. \ref{fig1}(a), we can calculate the
normalized scattering cross section (NSCS) based on the Mie scattering
theory \cite{bookEWPRS,bookASLSP}.
Detailed calculation shows that
\begin{equation}
\text{NSCS}=\sum\nolimits_{n=0}^{\infty }\delta _{n}\left\vert
s_{n}\right\vert ^{2},  \label{NSCS}
\end{equation}%
where%
\begin{equation}
s_{n}=-\frac{J_{n}^{\prime }\left( k_{0}R\right) t_{n}-J_{n}\left(
k_{0}R\right) J_{n}^{\prime }\left( kR\right) }{H_{n}^{\left( 1\right)
\prime }\left( k_{0}R\right) t_{n}-H_{n}^{\left( 1\right) }\left(
k_{0}R\right) J_{n}^{\prime }\left( kR\right) },  \label{sn}
\end{equation}%
$t_{n}=\sqrt{\varepsilon _{r}}J_{n}\left( kR\right) +i\sigma _{g}\eta
_{0}J_{n}^{\prime }\left( kR\right) $, $\delta _{n}=1$ for $n=0$ and $\delta
_{n}=2$ for $n\neq 0$, $k_{0}=\omega \sqrt{\varepsilon _{0}\mu _{0}}$ is the
wavenumber in free space, $k=k_{0}\sqrt{\varepsilon _{r}}$ is the wavenumber
in the dielectric cylinder, and $J_{n}$ and $H_{n}^{\left( 1\right) }$ are
the $n$-th order Bessel function of the first kind and Hankel function of
the first kind, respectively \cite{bookHMFFGMT}.
\textcolor[rgb]{0,0,0}{Since the degeneracy between $\left\vert s_{n}\right\vert $ and
$\left\vert s_{-n}\right\vert $ are considered in Eq. (\ref{NSCS}),
the NSCS has a single channel limit $\delta_{n}$ for the $n$th angular
momentum channel \cite{PRL105-013901}.}
While for the waveguide model in Fig. \ref{fig1} (b),
the dispersion relation of TM graphene plasmons is
\begin{equation}
\beta =k_{0}\sqrt{1+\left( \frac{2}{\sigma _{g,i}\eta _{0}}\right) ^{2}},
\label{dispersion relation}
\end{equation}%
where $\eta _{0}=\sqrt{\mu _{0}/\varepsilon _{0}}$ is the impendence of free
space, and $\sigma _{g,i}$ is the imaginary part of surface conductivity of
graphene \cite{science332-1291}. For simplicity, the real part of surface conductivity is omitted
in the calculation of dispersion relations in the whole paper to neglect the
optical loss of graphene.

Based on Eqs. (\ref{NSCS})-(\ref{dispersion relation}), Figs. \ref{fig1}(c)
and (d) show the NSCSs at different frequencies and the dispersion relation,
respectively. In Fig. \ref{fig1}(c), the dashed green line denotes the total
NSCS, while the solid black, red, and blue lines denote contributions from $%
n=0$, $n=1$, and $n=2$ scattering terms, respectively. The inset is the
enlarged figure. Clearly, the total NSCS exhibits two resonance peaks at $%
f=15.24$ THz and $f=21.52$ THz, which are caused by the resonances of $n=1$
and $n=2$ scattering terms, respectively. Note the NSCSs are expressed in
the common logarithmic form since the resonance peaks are very sharp. For
this scattering model, the two resonances correspond to the first and second
order Bohr conditions with the effective radius $R_{\text{eff}}=R$,
respectively.
\textcolor[rgb]{0,0,0}{The effective radius is determined because the
field intensity is maximum on
the graphene surface.}
As shown in Fig. \ref{fig1}(d), the two vertical dashed black
lines indicate the Bohr conditions of $\beta R_{\text{eff}}=1$ and $\beta R_{%
\text{eff}} = 2$, where the corresponding resonant frequencies are $f=15.27$
THz and $f=21.52$ THz, respectively. The calculation results from Bohr model
agree well with that from the scattering model, which implies that Bohr
model is applicable for the interpretation of scattering phenomena of
graphene-based structures.

\textcolor[rgb]{0,0,0}{Compared with the calculation result in Ref. \cite{sr5-12148}
where Bohr model is used to interpret the superscattering of a metal nanowire,
our results are more resonable due to the much tighter
confinement of graphene plasmons \cite{NL8-3370}. Thus Bohr model demonstrates
more advantages in designing ultra-compact graphene-based
superscatterers. }

\section{Dispersion engineering}

As discussed in the above section, the scattering cross sections of
plasmonic structures exhibit resonances where superscattering occurs at the
resonant frequencies. However, this kind of resonances are caused by the
resonance of a single scattering term, while the contributions from different
scattering terms can be overlapped to further enhance the scattering cross
sections \cite{PRL105-013901,APL98-043101}. Due to the complexity of
scattering models, the genetic algorithm has been used to optimize the
superscattering of light where different resonance peaks are overlapped \cite%
{APL105-011109}. However, it still lacks an intuitive method to design the
superscatterers.

Based on the Bohr model, the scattering model is related to its equivalent
planar waveguide model intuitively. Thus, it is possible to enhance the
scattering cross sections by engineering the dispersion relations of the
equivalent planar waveguides. According to Eqs. (\ref{Bohr condition})-(\ref%
{reduced Bohr condition}), different orders of Bohr conditions must be
satisfied at the same frequency to overlap the resonance peaks. This
requirement can be fulfilled in a single dispersion curve or multiple
dispersion curves.

\textcolor[rgb]{0,0,0}{In Refs. \cite{PRL105-013901,APL98-043101},
different orders of Bohr
conditions are satisfied in a single dispersion curve. The
metal-dielectric-metal-dielectric-air planar waveguide supports a flat
dispersion curve with a proper choice of the thickness of the dielectric
layers. Thus, different scattering terms are overlapped and the scattering
cross section is enhanced. Alternatively, in this paper we enhance the
scattering cross sections of plasmonic structures from a different approach
where Bohr conditions with different orders are fulfilled in multiple
dispersion curves.}

\begin{figure}[tbp]
\centering
\vspace{0.1cm} \includegraphics[width=8.5cm]{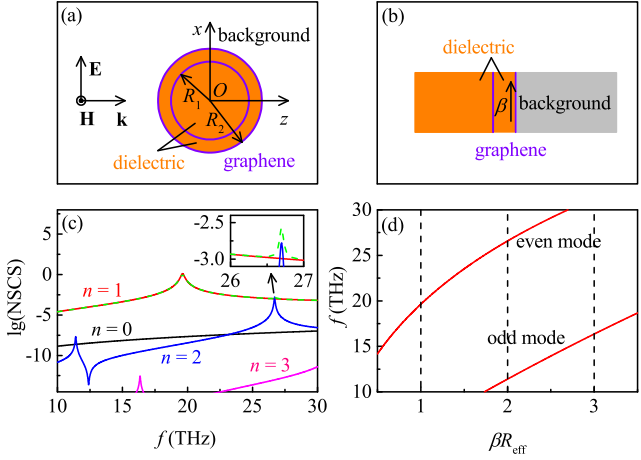} \vspace{-0.0cm}
\caption{(Color Online) (a) Cross-sectional view of the layered
dielectric-graphene-dielectric-graphene-background cylindrical structure.
The dielectric layers are denoted by the orange areas, and the graphene
layer is denoted by the violet area. A $x$-polarized plane wave is incident
from background onto the structure. The radii of the inner and outer
dielectric layers are $R_{1}$ and $R_{2}$, respectively. (b) Structure of
the equivalent planar waveguide model, where the gap thickness between two
graphene layers is $d=R_{2}-R_{1}$. The graphene plasmons propagate along
the graphene surface in the direction indicated by an arrow, and $\protect%
\beta$ is the corresponding propagation constant. (c) The normalized
scattering cross sections (NSCSs) at different frequencies for the structure
shown in (a), where the dashed green line denotes the total NSCS, while the
solid black, red, blue, and magenta lines denote contributions from $n=0$, $%
n=1$, $n=2$, and $n=3$ scattering terms, respectively. The inset is the
enlarged figure. Note the NSCSs are expressed in the common logarithmic
form. (d) Dispersion relation of graphene plasmons for the waveguide
structure shown in (b). The three vertical dashed black lines indicate the
first ($\protect\beta R_{\text{eff}}=1$), second ($\protect\beta R_{\text{eff%
}}=2$), and third ($\protect\beta R_{\text{eff}}=3$) order Bohr conditions,
respectively. The parameters are $\protect\varepsilon_{1}=\protect\varepsilon%
_{2}=\protect\varepsilon_{b}=1$, $\protect\mu_{1}=\protect\mu_{2}=\protect\mu%
_{b}=1$, $R_{1}=250$ nm, $R_{2}=300$ nm, $d=R_{2}-R_{1}=50$ nm, $\protect\mu%
_c=0.35$ eV, $\textcolor[rgb]{0,0,0}{\mu}=$ 85 600 $\text{cm}^{2}/(\text{V}\cdot\text{s})$,
and $T=300$ K for (c) and (d). }
\label{fig2}
\end{figure}

For simplicity, we consider a layered
dielectric-graphene-dielectric-graphene-background cylindrical structure, as
shown in Fig. \ref{fig2}(a). A $x$-polarized plane wave is incident normally
from the background onto the structure. The background is a dielectric
medium with the relative permittivity $\varepsilon_{b}$ and relative
permeability $\mu_{b}=1$. The relative permittivities of the inner and outer
dielectric layers are $\varepsilon _{1} $ and $\varepsilon _{2}$,
respectively, and the relative permeabilities are $\mu _{1}=\mu _{2}=1$. The
radii of the inner and outer cylindrical graphene layers are $R_{1}$ and $%
R_{2}$, respectively. Besides, the parameters of the graphene layer are $\mu
_{c}=0.35$ eV, $\textcolor[rgb]{0,0,0}{\mu}=$ 85 600
$\text{cm}^{2}/(\text{V}\cdot\text{s})$,
and $T=300$ K. Similarly, for the
scattering model in Fig. \ref{fig2}(a), the corresponding equivalent planar
waveguide structure is shown in Fig. \ref{fig2}(b), where $d=R_{2}-R_{1} $
is the gap thickness between two graphene layers.

For the scattering model in Fig. \ref{fig2}(a), the normalized scattering
cross section (NSCS) is
\begin{equation}
\text{NSCS}=\sum\nolimits_{n=0}^{\infty }\delta _{n}\left\vert
s_{n}\right\vert ^{2},  \label{NSCS1}
\end{equation}%
where%
\begin{align}
s_{n}& =-\frac{J_{n}^{\prime }\left( k_{b}R_{2}\right) q_{n}-\sqrt{%
\varepsilon _{b}}J_{n}\left( k_{b}R_{2}\right) r_{n}}{H_{n}^{\left( 1\right)
\prime }\left( k_{b}R_{2}\right) q_{n}-\sqrt{\varepsilon _{b}}H_{n}^{\left(
1\right) }\left( k_{b}R_{2}\right) r_{n}},  \label{sn1} \\
q_{n}& =\sqrt{\varepsilon _{2}}\left[ J_{n}\left( k_{2}R_{2}\right)
+t_{n}H_{n}^{\left( 1\right) }\left( k_{2}R_{2}\right) \right] +i\sigma
_{g}\eta _{0}r_{n},  \label{qn} \\
r_{n}& =J_{n}^{\prime }\left( k_{2}R_{2}\right) +t_{n}H_{n}^{\left( 1\right)
\prime }\left( k_{2}R_{2}\right) ,  \label{rn} \\
t_{n}& =-\frac{J_{n}^{\prime }\left( k_{2}R_{1}\right) p_{n}-\sqrt{%
\varepsilon _{2}}J_{n}\left( k_{2}R_{1}\right) J_{n}^{\prime }\left(
k_{1}R_{1}\right) }{H_{n}^{\left( 1\right) \prime }\left( k_{2}R_{1}\right)
p_{n}-\sqrt{\varepsilon _{2}}H_{n}^{\left( 1\right) }\left(
k_{2}R_{1}\right) J_{n}^{\prime }\left( k_{1}R_{1}\right) },  \label{tn} \\
p_{n}& =\sqrt{\varepsilon _{1}}J_{n}\left( k_{1}R_{1}\right) +i\sigma
_{g}\eta _{0}J_{n}^{\prime }\left( k_{1}R_{1}\right) ,  \label{pn}
\end{align}%
and $k_{1}=k_{0}\sqrt{\varepsilon _{1}}$, $k_{2}=k_{0}\sqrt{\varepsilon _{2}}
$, and $k_{b}=k_{0}\sqrt{\varepsilon _{b}}$ are wavenumbers in the inner and
outer dielectric layers, and the background, respectively \cite{bookEWPRS,bookASLSP}.
While for the
waveguide model in Fig. \ref{fig2}(b), the dispersion relation of TM
graphene plasmons is
\begin{align}
& e^{-2k_{2}d}  \notag \\
=& \frac{\frac{k_{2}}{\varepsilon _{2}}\left( 1+i\sigma _{g}\frac{k_{b}}{%
\omega \varepsilon _{0}\varepsilon _{b}}\right) +\frac{k_{b}}{\varepsilon
_{b}}}{\frac{k_{2}}{\varepsilon _{2}}\left( 1+i\sigma _{g}\frac{k_{b}}{%
\omega \varepsilon _{0}\varepsilon _{b}}\right) -\frac{k_{b}}{\varepsilon
_{b}}}\frac{\frac{k_{2}}{\varepsilon _{2}}\left( 1+i\sigma _{g}\frac{k_{1}}{%
\omega \varepsilon _{0}\varepsilon _{1}}\right) +\frac{k_{1}}{\varepsilon
_{1}}}{\frac{k_{2}}{\varepsilon _{2}}\left( 1+i\sigma _{g}\frac{k_{1}}{%
\omega \varepsilon _{0}\varepsilon _{1}}\right) -\frac{k_{1}}{\varepsilon
_{1}}}.  \label{dispersion relation new}
\end{align}%
Speically, when $\varepsilon _{1}=\varepsilon _{b}$, this waveguide is a
symmetric double-channel graphene plasmon waveguide, and the dispersion
relation reduces to
\begin{equation}
\tanh \left( \frac{k_{2}d}{2}\right) =-\frac{k_{1}\varepsilon _{2}}{%
k_{2}\varepsilon _{1}\left( 1-\sigma _{g,i}\frac{k_{1}}{\omega \varepsilon
_{0}\varepsilon _{1}}\right) }  \label{dispersion relation odd}
\end{equation}%
for the odd mode, and%
\begin{equation}
\tanh \left( \frac{k_{2}d}{2}\right) =-\frac{k_{2}\varepsilon _{1}\left(
1-\sigma _{g,i}\frac{k_{1}}{\omega \varepsilon _{0}\varepsilon _{1}}\right)
}{k_{1}\varepsilon _{2}}  \label{dispersion relation enven}
\end{equation}%
for the even mode \cite{bookmaier}. Note the odd and even modes are defined according to the
parity of the tangential electric component $E_{x}$.

Based on Eqs. (\ref{NSCS1})-(\ref{dispersion relation enven}), we can tune
the positions of the resonance peaks from different scattering terms by dispersion engineering.
However, before the demonstration of dispersion engineering, first we need
to validate the applicability of Bohr model when multiple dispersion curves
exist simultaneously. We consider the simplest case with $\varepsilon
_{1}=\varepsilon _{2}=\varepsilon _{b}=1$, $R_{1}=250$ nm, $R_{2}=300$ nm,
and $d=R_{2}-R_{1}=50$ nm.
Under these parameters, Figs. \ref{fig2}(c) and
(d) show the NSCSs at different frequencies and the dispersion relation,
respectively. In Fig. \ref{fig2}(c), the dashed green line denotes the total
NSCS, while the solid black, red, blue, and magenta lines denote
contributions from $n=0$, $n=1$, $n=2$, and $n=3$ scattering terms,
respectively. The inset is the enlarged figure. Clearly, the total NSCS
exhibits four resonance peaks, where the peak at $f=19.62$ THz is caused by
the resonance of $n=1$ scattering term, the peaks at $f=11.40$ THz and $%
f=26.70$ are caused by the resonances of $n=2$ scattering term, and the peak
at $f=16.35$ THz is caused by the resonance of $n=3$ scattering term,
respectively. Note the resonance peaks of total NSCS at $f=11.40$ THz and $%
f=16.35$ THz are not pronounced since the contributions from $n=2$ and $n=3$
scattering terms at the corresponding resonant frequencies are small. For
this scattering model, these resonances are related to the Bohr conditions
with the effective radius $R_{\text{eff}}=(R_{1}+R_{2})/2$.
\textcolor[rgb]{0,0,0}{This effective radius is determined because
it corresponds to the bisector of the two graphene layers.}
As shown in Fig. %
\ref{fig2}(d), the first order Bohr condition $\beta R_{\text{eff}}=1$ at $%
f=19.63$ THz corresponds to the resonance of $n=1$ scattering term, the
third order Bohr condition $\beta R_{\text{eff}}=3$ at $f=16.35$ THz
corresponds to the resonance of $n=3$ scattering term, and the second order
Bohr conditions $\beta R_{\text{eff}}=2$ at $f=26.58$ THz for the even mode
and $f=11.40$ THz for the odd mode correspond to the resonances of $n=2$
scattering term at $f=11.40$ THz and $f=26.70$ THz, respectively. The
calculation results from Bohr model agree well with that from the scattering
model, which implies that Bohr model is still applicable when multiple
dispersion curves exist simultaneously.

\begin{figure}[tbp]
\centering
\vspace{0.1cm} \includegraphics[width=8.5cm]{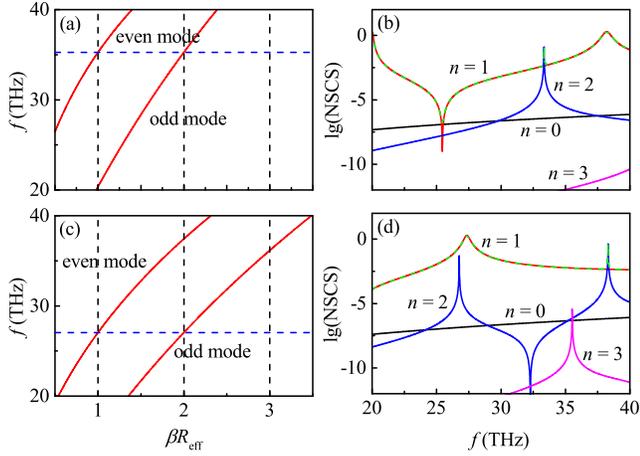} \vspace{-0.0cm}
\caption{(Color Online) Left panel: Dispersion relations of graphene
plasmons for the waveguide structure shown in Fig. \protect\ref{fig2}(b)
with (a) $\protect\varepsilon_{1}=\protect\varepsilon_{2}=\protect\varepsilon%
_{b}=1$ and $d=138.20$ nm, and (c) $\protect\varepsilon_{1}=\protect%
\varepsilon_{b}=3$, $\protect\varepsilon_{2}=1$, and $d=44.38$ nm. The three
vertical dashed black lines in each figure indicate the first ($\protect%
\beta R_{\text{eff}}=1$), second ($\protect\beta R_{\text{eff}}=2$), and
third ($\protect\beta R_{\text{eff}}=3$) order Bohr conditions,
respectively. The resonant frequencies where the resonant peaks are
overlapped (indicated by the dashed blue lines) are $f=35.26$ THz for (a)
and $f=27.04$ THz for (c), respectively. Right panel: The normalized
scattering cross sections (NSCSs) at different frequencies for the structure
shown in Fig. \protect\ref{fig2}(a) with (b) $\protect\varepsilon_{1}=%
\protect\varepsilon_{2}=\protect\varepsilon_{b}=1$, $R_{1}=131.54$ nm, and $%
R_{2}=269.74$ nm, and (d) $\protect\varepsilon_{1}=\protect\varepsilon_{b}=3$%
, $\protect\varepsilon_{2}=1$, $R_{1}=122.06$ nm, and $R_{2}=166.43$ nm. In
each figure, the dashed green line denotes the total NSCS, while the solid
black, red, blue, and magenta lines denote contributions from $n=0$, $n=1$, $%
n=2$, and $n=3$ scattering terms, respectively. Note the NSCSs are expressed
in the common logarithmic form. The other parameters are $\mu
_{c}=1$ eV, $\textcolor[rgb]{0,0,0}{\mu}=$ 230 000
$\text{cm}^{2}/(\text{V}\cdot\text{s})$,
and $T=300$ K.}
\label{fig3}
\end{figure}

The idea of overlapping the resonance peaks by dispersion engineering is
very straightforward. From Fig. \ref{fig2}(d), the dispersion curves of the
even mode and odd mode are dependent on the gap thickness $d$. By delicately
tuning the gap thickness, the first order Bohr condition for the even mode
and the second order Bohr condition for the odd mode can be fulfilled at the
same frequency.
In the following calculations, the parameters of the graphene layer are $\mu
_{c}=1$ eV, $\textcolor[rgb]{0,0,0}{\mu}=$ 230 000
$\text{cm}^{2}/(\text{V}\cdot\text{s})$,
and $T=300$ K to reduce the optical loss of graphene.
As shown in Fig. \ref{fig3}(a), when $%
\varepsilon_{1}=\varepsilon_{2}=\varepsilon_{b}=1$ and the gap thickness $d
= 138.20$ nm, the first order Bohr condition for the even mode and the
second order Bohr condition for the odd mode are fulfilled at $f = 35.26$
THz. Then we can obtain the radii of the inner and outer graphene layers
according to the Bohr condition, namely $R_{1}=131.54$ nm and $%
R_{2}=269.74$ nm. However, under these parameters the resonance peaks from $n=1$
and $n=2$ are not overlapped at $f = 35.26$ THz, as shown in Fig. \ref{fig3}%
(b). This deviation is due to the invalidation of Bohr model. As shown in
Fig. \ref{fig2}(a)-(b), Bohr model relates the scattering model to the
waveguide model based on the assumption that the gap thickness $d$ is
smaller than the radius of the inner graphene layer. Under this assumption,
the excited whispering-gallery-like modes can be approximated by the
plasmonic modes in the equivalent one dimensional planar waveguide. For Fig. %
\ref{fig3}(b), since the gap thickness is nearly equal to the value of $%
R_{1} $, the resonance peaks deviate from the resonant frequency predicted
by Bohr model.

\begin{figure}[tbp]
\centering
\vspace{0.1cm} \includegraphics[width=8.5cm]{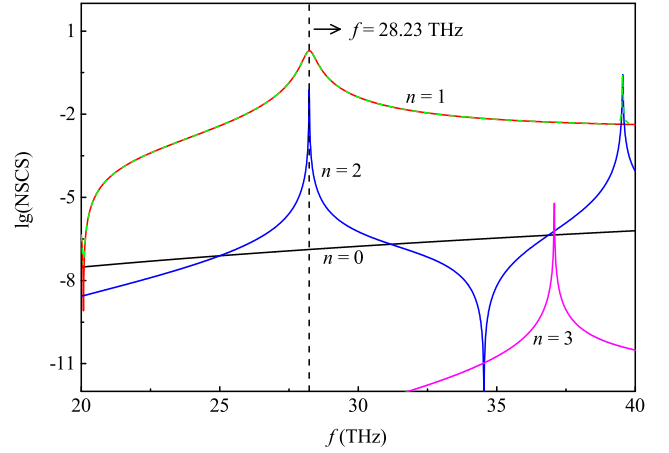} \vspace{-0.0cm}
\caption{(Color Online) The normalized scattering cross sections (NSCSs) at
different frequencies for the structure shown in Fig. \protect\ref{fig2}(a),
where the dashed green line denotes the total NSCS, while the solid black,
red, blue, and magenta lines denote contributions from $n=0$, $n=1$, $n=2$, and
$n=3$ scattering terms, respectively. Note the NSCSs are expressed in the
common logarithmic form. The parameters are $\protect\varepsilon_{1}=\protect%
\varepsilon_{b}=3$, $\protect\varepsilon_{2}=1$, $\protect\mu_{1}=\protect\mu%
_{2}=\protect\mu_{b}=1$, $R_{1} = 112.47$ nm, $R_{2} = 160.24$ nm, $\protect%
\mu_c=1$ eV, $\textcolor[rgb]{0,0,0}{\mu}=$ 230 000
$\text{cm}^{2}/(\text{V}\cdot\text{s})$,
and $T=300$ K. As indicted by the dashed
black line, the resonant frequency where the resonant peaks are overlapped
is $f=28.23$ THz.}
\label{fig4}
\end{figure}

In order to fulfill the Bohr condition, we let $\varepsilon_{1}=%
\varepsilon_{b}=3$, $\varepsilon_{2}=1$, and $d=44.38$ nm. Note the
background is not air for demonstration purpose. Similarly, as shown in Fig. %
\ref{fig3}(c), the first order Bohr condition for the even mode and the
second order Bohr condition for the odd mode are are fulfilled at $f = 27.04$
THz, and we can obtain $R_{1}=122.06$ nm and $R_{2}=166.43$ nm. Under
these parameters, the resonance peaks from $n=1$ and $n=2$ are overlapped at
$f = 27.04$ THz with a small deviation, as shown in Fig. \ref{fig3}(d).
Clearly, Bohr condition is valid since $d<R_{1}$ in this case. The above
result can be optimized further using the simplex search method \cite{simplex}, where $%
R_{1} = 122.06$ nm and $d=44.38$ nm are set as the initial estimates. This
optimization method can find the local optimum values starting at the
initial estimates. Fig. \ref{fig4} show the normalized scattering cross
sections (NSCSs) at different frequencies under the optimized values of $%
R_{1} = 112.47$ nm and $R_{2} =R_{1}+d= 160.24$ nm. Note the resonant
frequency slightly changes to $f = 28.23$ THz. Thus the resonant peaks are
overlapped by dispersion engineering based on the Bohr model.
Note that the total normalized scattering cross section at $f = 28.23$ THz is 2.02,
which exceeds the single channel limit of a single scattering term \cite{PRL105-013901}.
This implies
that it is possible to design the ultra-compact graphene-based superscatterers from an
intuitive view.

\section{Design of Superscatterers}

In the above section, the resonance peaks are overlapped based on the
layered dielectric-graphene-dielectric-graphene-background cylindrical structures.
For demonstration purpose, the background is chosen as the dielectric medium
with $\varepsilon_{b}=3$. Considering the practical applications, the
background is the air with $\varepsilon_{b}=1$. Thus it is necessary
to take this restriction into account when designing the superscatterers.
In this section, based on the structure shown in Fig. \ref{fig2}(a) with $\varepsilon_{b}=1$,
we design an ultra-compact graphene-based superscatterer by dispersion engineering.
\begin{figure}[tbp]
\centering
\vspace{0.1cm} \centerline{\includegraphics[width=8.5cm]{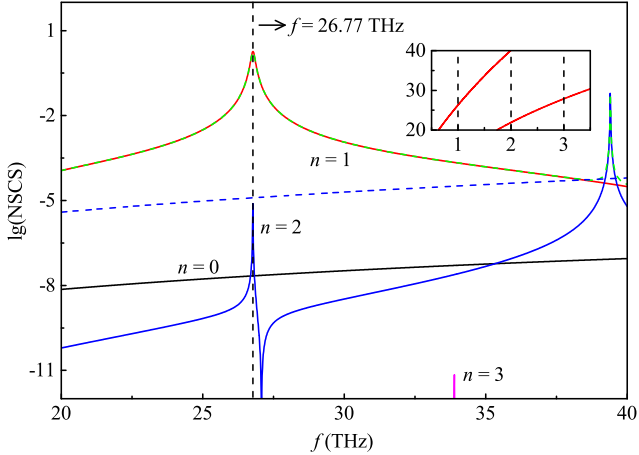}} \vspace{%
-0.0cm}
\caption{(Color Online) The normalized scattering cross sections (NSCSs) at
different frequencies for the designed superscatterer,
where the dashed green line denotes the total NSCS, while the solid black,
red, blue, and magenta lines denote contributions from $n=0$, $n=1$, $n=2$, and
$n=3$ scattering terms, respectively.
\textcolor[rgb]{0,0,0}{For comparison, the dashed blue line shows the
total NSCS of a dielectric
cylinder with $R= 118.88$ nm.}
Note the NSCSs are expressed in the
common logarithmic form. The parameters are $\protect\varepsilon_{1}=6$,
$\protect\varepsilon_{2}=1.1$,
$\varepsilon_{b}=1$,  $\protect\mu_{1}=\protect\mu%
_{2}=\protect\mu_{b}=1$, $R_{1} =  118.88$ nm, $R_{2} =  193.00$ nm, $\protect%
\mu_c=1$ eV, $\textcolor[rgb]{0,0,0}{\mu}=$ 230 000
$\text{cm}^{2}/(\text{V}\cdot\text{s})$,
and $T=300$ K. As indicted by the dashed
black line, the resonant frequency where the resonant peaks are overlapped
is $f=26.77$ THz. The inset shows the dispersion relation of graphene plasmons
in the equivalent planar waveguide.}
\label{fig5}
\end{figure}

Following the similar procedure, we let $\varepsilon_{1}=6$, $\varepsilon_{2}=1.1$,
and $\varepsilon_{b}=1$, and tune the gap thickness $d$ between two
graphene layers. When $d=63.48$ nm, the first order Bohr condition is fulfilled
at $f=26.30$ THz, the second order Bohr condition is fulfilled at $f=21.90$ THz,
and the radii of the inner graphene layer is $R_{1}=140.623$ nm. Using the
simplex search method, we obtain the optimized values of $R_{1}= 118.88$ nm
and $R_{2}= R_{1}+ d = 193.00$ nm. Fig. \ref{fig5} shows
the normalized scattering cross sections (NSCSs) at different
frequencies, where the resonance peaks are overlapped at the resonant frequency
$f = 26.77$ THz. Note that the normalized scattering
cross section of a dielectric
cylinder with $R= 118.88$ nm is $1.25\times10^{-5}$ at $f = 26.77$ THz,
\textcolor[rgb]{0,0,0}{as shown in Fig. \ref{fig5}.}
Although the total normalized scattering cross section of the superscatterer
shown in Fig. \ref{fig5}
is 1.82 which is still under the single channel limit, the scattering cross section
contributed by the overlapped resonance peaks
has enhanced for five orders of magnitude.
Meanwhile, the radius of the ultra-compact superscatterer is only $0.017\lambda_{0}$,
where $\lambda_{0}$ is the incident wavelength.
Besides, the optimized values of the superscatterer can be easily obtained
by the local optimization algorithm.
Compared with the global optimization algorithms, the method of
dispersion engineering provides an intuitive way to design the
ultra-compact graphene-based superscatterers.

\begin{figure}[tbp]
\centering
\vspace{0.1cm} \centerline{\includegraphics[width=8.5cm]{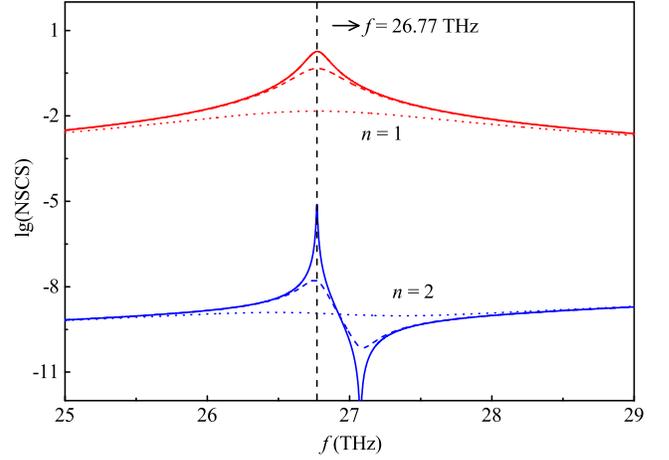}} \vspace{%
-0.0cm}
\caption{\textcolor[rgb]{0,0,0}{(Color Online) The normalized
scattering cross sections (NSCSs) at
different frequencies for the designed superscatterer under different values
of carrier mobility of graphene,
where the red and blue lines denote contributions from $n=1$ and $n=2$
scattering terms, respectively.
For simplicity, the curves for $n=0$ and $n=3$
are omitted. Note the NSCSs are expressed in the
common logarithmic form. The parameters are $\protect\varepsilon_{1}=6$,
$\protect\varepsilon_{2}=1.1$,
$\varepsilon_{b}=1$,  $\protect\mu_{1}=\protect\mu%
_{2}=\protect\mu_{b}=1$, $R_{1} =  118.88$ nm, $R_{2} =  193.00$ nm, $\protect%
\mu_c=1$ eV, $T=300$ K, $\mu=$ 230 000 $\text{cm}^{2}/(\text{V}\cdot\text{s})$
for the solid lines, $\mu=$ 10 000 $\text{cm}^{2}/(\text{V}\cdot\text{s})$
for the dashed lines, and $\mu=$ 1 000 $\text{cm}^{2}/(\text{V}\cdot\text{s})$
for the dotted lines. As indicted by the dashed
black line, the resonant frequency where the resonant peaks are overlapped
is $f=26.77$ THz.}}
\label{fig6}
\end{figure}

\textcolor[rgb]{0,0,0}{Since the plasmonic field is highly localized
on the graphene surface,
the superscattering phenomenon is sensitive to the optical loss of graphene.
Fig. \ref{fig6} shows the normalized scattering cross sections (NSCSs) at
different frequencies for the designed superscatterer under different values of
carrier mobility of graphene. When the carrier mobility decreases,
namely the optical loss of graphene increases, the resonant
frequency is fixed and the NSCSs from both the $n=1$ and $n=2$ scattering
terms decrease. Specially, the higher order modes are more susceptible to the
optical loss of graphene \cite{nanotechnology,PRL105-013901}.
This implies that, graphene monolayers with high carrier mobilities
are necessary to improve the performance of our ultra-compact graphene-based
superscatterers.}

\textcolor[rgb]{0,0,0}{Finally, we compare our superscatterers designed by
dispersion engineering with those designed using transformation optics.
Compared with dispersion engineering, transformation optics is a more
general method that can be used to control the scattering of a given
object freely. Apart from enhancing the scattering by superscatterers,
the scattering can
also be suppressed to realize cloaks \cite{PRL102-093901,APL97-133501}.
Besides, transformation optics
can also be used to design electrically-small antennas \cite{APL95-193506,NJP12-033047}.
These devices are hard to realize by dispersion engineering from an intuitive
way. However, superscatterers designed by transformation optics are difficult
to implement in experiments since the electromagnetic covers are made of
anisotropic and inhomogeneous materials. In contrast, the method of
dispersion engineering is more intuitive. More importantly, the superscatterers
designed by dispersion engineering are more easier to implement
where only isotropic and homogeneous materials are used.}

\section{Conclusions}

In conclusion, based on the validation of Bohr model, we show that
the method of dispersion engineering can be introduced to the design
of ultra-compact graphene-based superscatterers.
Since Bohr conditions with different orders
are fulfilled in multiple dispersion curves at the same resonant
frequency,
the resonance peaks from the first
and second order scattering terms are overlapped in the deep-subwavelength
scale to further enhance
the scattering cross sections by five orders of magnitude.
Compared with the global optimization
algorithms, our method is more intuitive in physics.
Our work will provide theoretical guidance for the
design of superscatterers and other scattering based devices, which have great
potential applications in plasmonics.


%
%
%
\begin{IEEEbiography}[{\includegraphics[width=1in,height=1.25in,clip,keepaspectratio]{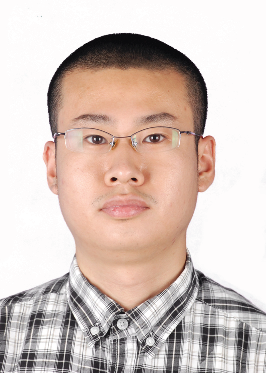}}]{Rujiang Li}
received the B.S. degree in physics (National Scientific Base for Talented Persons) and M.S. degree in theoretical
physics from Shanxi University, Taiyuan, China, in 2011 and 2014, respectively. He is currently working toward the
Ph.D. degree in electronic science and technology at Zhejiang University, Hangzhou, China.

His current research interests include scattering by subwavelength structures, nonlinear optics and
nonlinear plasmonics, and graphene.
\end{IEEEbiography}
\begin{IEEEbiography}[{\includegraphics[width=1in,height=1.25in,clip,keepaspectratio]{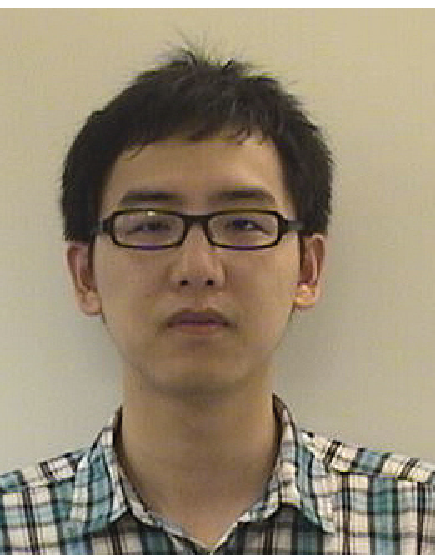}}]{Bin Zheng}
received the B.S. degree from Ningbo University, Ningbo, China, in 2010; and the Ph.D. degree from Zhejiang
University, Hangzhou, China, in 2015. He is currently a postdoctoral researcher in College of Information
Science \& Electronic Engineering, Zhejiang University, China, since 2015.
\end{IEEEbiography}
\begin{IEEEbiography}[{\includegraphics[width=1in,height=1.25in,clip,keepaspectratio]{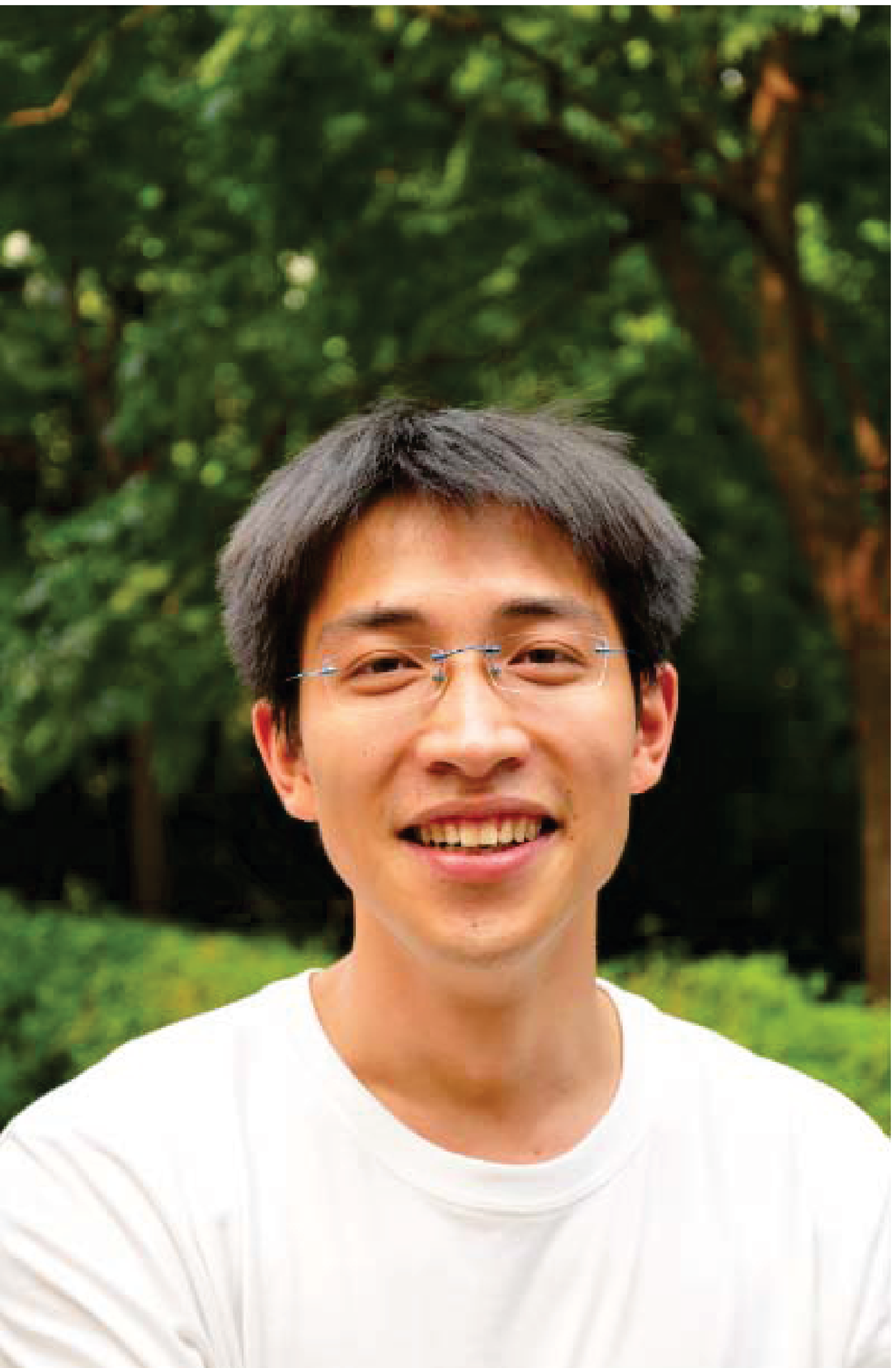}}]{Xiao Lin}
received the B.S. degree
in optical science and engineering
from Zhejiang University (ZJU), Hangzhou,
China in 2011. He is currently working toward the Ph.D. degree at the college of information
science and electronic engineering in ZJU. During his Ph.D. studies, he was a visiting
student at the Nanyang Technological University (NTU), Singapore and the Massachusetts Institute
of Technology (MIT), USA, respectively.
His current interest includes 2D materials, surface plasmons, and electromagnetic radiation.
\end{IEEEbiography}
\begin{IEEEbiography}[{\includegraphics[width=1in,height=1.25in,clip,keepaspectratio]{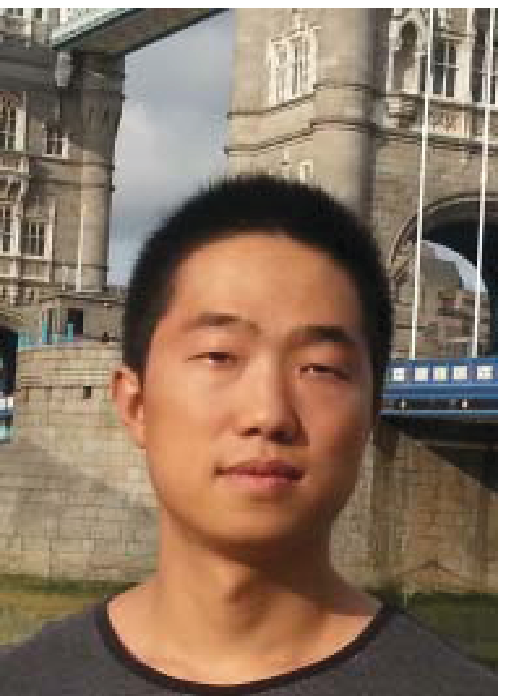}}]{Ran Hao}
received his Ph.D. degree in physics from University Paris XI, France in December, 2010. And he received his
second Ph.D. degree in Photonics from Wuhan National Laboratory for optoelectronics, Huazhong University of
Science \& Technology, China, in 2011. He is currently an associate professor in College
of information science \& electronic engineering, Zhejiang University, China. He has won the Distinguished Young
Scholar award in 2011, the Excellent Young Faculty Awards Program in 2012 and 2013, and the Young Scientist Award
from the General Assembly and Scientific Symposium of International Union of Radio Science in 2014.
His current research interests include plasmonics, nanophotonics, photonic crystals,
and graphene photonics. Dr. Hao is a senior member of IEEE photonic society and member of
the OSA, SPIE and COS.
\end{IEEEbiography}
\begin{IEEEbiography}[{\includegraphics[width=1in,height=1.25in,clip,keepaspectratio]{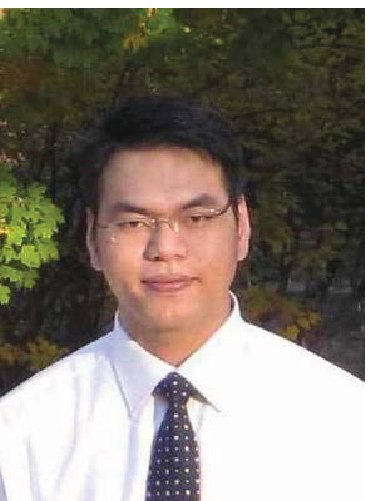}}]{Shisheng Lin}
received his Ph. D. degree in Materials Physics \& Chemistry under the joint education
of Zhejiang University and Georgia Institute of Technology, in 2010. He is currently an Associate Professor in
Zhejiang University. Prof. Lin focused on doping physics in ZnO and its based optoelectronic devices,
fabrication and Raman physics of graphene, graphene based optoelectronic devices, 2D materials
based metamaterials and energy harvesting devices.
Prof. Lin has published more than 50 international
refereed journal papers with over 700 times citations.
\end{IEEEbiography}
\begin{IEEEbiography}[{\includegraphics[width=1in,height=1.25in,clip,keepaspectratio]{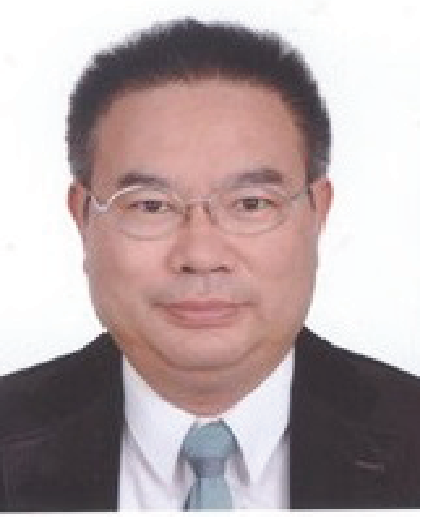}}]{Wenyan Yin}
(M'99-SM'01-F'13) received the M.S. degree
from Xidian University, Xi'an, China, in 1989, and the Ph.D. degree
in electrical engineering
from Xi'an Jiao Tong University, Xi'an, China, in 1994. From 1993 to 1996, he was an Associate Professor
in Northwestern Polytechnic University, Xi'an, China. From 1996 to 1998, he
was a Research Fellow
in Duisburg University, Duisburg, Germany.
Since
1998, he has been with the National University of Singapore,
Singpore, as a Research Scientist and the Project Leader.
Since
2005, he has
been a Professor
in Shanghai Jiao Tong University, Shanghai, China, where he is currently an Adjunct
Ph.D. Candidate Supervisor.
In
2009, he joined Zhejiang University,
China, as a ¡°Qiu Shi¡± Distinguished Professor.
He has written more than 200 international journal papers. 
His main research interests are passive and active RF and millimeter-wave device and circuit modeling, ultra-wideband
interconnects and signal integrity, nanoelectronics, electromagnetic compatibility (EMC) and electromagnetic protection
of communication platforms, and computational multi-physics and its application.

Dr. Yin is an associate editor for the IEEE Transactions on Components, Packaging, and Manufacturing Technology.
From 2011 to 2012, he was an IEEE EMC Society Distinguished Lecturer. Since 2013, he has been the IEEE EMC Society
Chapter Chair.
Since 2011, he has been an associate editor of the International Journal of Electronic Networks,
Devices, and Fields. He is also an Editorial Board member of International Journal of RF and Microwave Computer-Aided
Engineering.
He was the recipient of the
the National Technology Invention Award of the Second Class from the Chinese Government in 2008.
\end{IEEEbiography}
\begin{IEEEbiography}[{\includegraphics[width=1in,height=1.25in,clip,keepaspectratio]{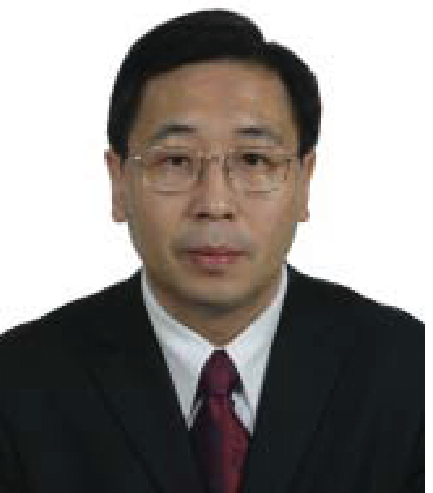}}]{Erping Li}
(S'91-M'92-SM'01-F'08) received the Ph.D. Degree in electrical engineering from Sheffield Hallam University,
Sheffield, U.K, in 1992. From 1993 to 1999, he was a Senior Research Fellow, Principal Research Engineer,
Associate Professor and the Technical Director at the Singapore Research Institute and Industry.
In 2000, he joined the Singapore National Research Institute of High Performance Computing as a
Principal Scientist and Director of the Electronic and Photonics Dept. He also holds the Distinguished Professor
at Zhejiang University. He authored or co-authored over 400 papers published in the referred international journals
and conferences, authored two books. His research interests include electrical modeling and design of micro/nano-scale
integrated circuits, 3D electronic package integration and nano-plasmonic technology.

Dr. Li is a Fellow of IEEE, and a Fellow of MIT Electromagnetics Academy, USA. He received numerous awards
including the IEEE EMC Richard Stoddard Award for outstanding performance, He has served as an Associate
Editor for number of IEEE Transactions and Letters. He has served as a General Chair and Technical Chair,
for many international conferences. He was the founding General Chair for the 2008, 2010 and 2012
Asia-Pacific EMC Symposium. He has been invited to give numerous invited talks and plenary speeches at
various international conferences and forums.
\end{IEEEbiography}
\begin{IEEEbiography}[{\includegraphics[width=1in,height=1.25in,clip,keepaspectratio]{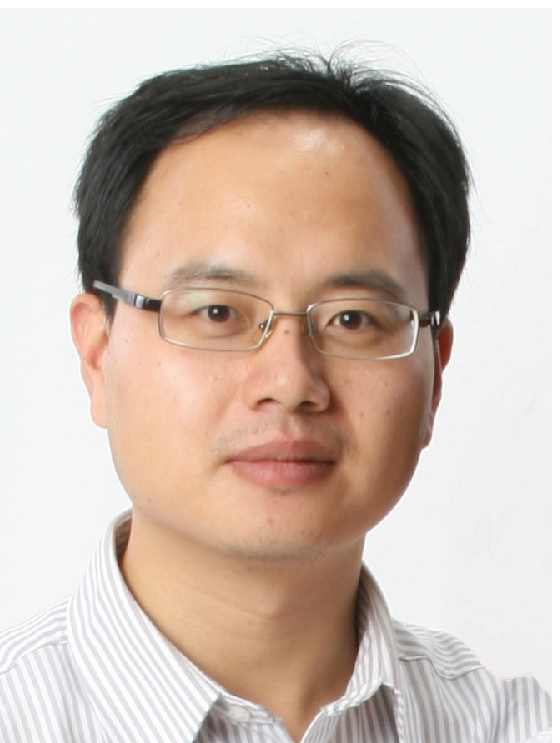}}]{Hongsheng Chen}
is a Chang Jiang Scholar Distinguished Professor in the Electromagnetics Academy at Zhejiang University in Hangzhou,
Zhejiang, China. He received the B.S. degree in 2000, and Ph.D. degree in 2005, from Zhejiang University, both in
electrical engineering.

In 2005, Chen became an Assistant Professor at Zhejiang University; in 2007 an Associate Professor; and in 2011 a
Full Professor. In 2014, he was honored with the distinguished "Chang Jiang Scholar" professorship by the Chinese
Ministry of Education. He was a Visiting Scientist (2006-2008), and a Visiting Professor (2013-2014) with the Research
Laboratory of Electronics at Massachusetts Institute of Technology, USA. His current research interests are in the areas
of metamaterials, antennas, invisibility cloaking, transformation optics, graphene, and theoretical and numerical methods
of electromagnetics. He is the coauthor of more than 130 international refereed journal papers. His works have been
highlighted by many scientific magazines and public media, including Nature, Scientific American, MIT Technology Review,
Physorg, and so on. He serves as a regular reviewer of many international journals on electromagnetics, physics, optics,
and electrical engineering. He serves on the Topical Editor of Journal of Optics, the Editorial Board of the Nature's
Scientific Reports, and Progress in Electromagnetics Research.

Dr. Chen was a recipient of National Excellent Doctoral Dissertation Award in China (2008), the Zhejiang Provincial
Outstanding Youth Foundation (2008), the National Youth Top-notch Talent Support Program in China (2012), the New Century
Excellent Talents in University of China (2012), and the National Science Foundation for Excellent Young Scholars of
China (2013). His research work on invisibility cloak was selected in Science Development Report as one of the representative
achievements of Chinese Scientists in 2007.
\end{IEEEbiography}
%
%
%

%

\end{document}